\documentclass[12pt]{article}
\usepackage{epsfig}
\usepackage{graphicx}
\usepackage{amssymb}
\usepackage{amsmath}
\usepackage{amsfonts}
\usepackage{mathrsfs}

\newcommand{\R}{\mathbb{R}}
\newcommand{\C}{\mathbb{C}}

\newcommand{\fa}{\mathfrak{a}}

\newcommand{\fn}{\mathfrak{n}}

\newcommand{\fz}{\mathfrak{z}}

\newcommand{\be}{\begin{equation}}
\newcommand{\ee}{\end{equation}}
\newcommand{\bea}{\begin{eqnarray}}
\newcommand{\eea}{\end{eqnarray}}
\newcommand{\nn}{\nonumber}

\newcommand{\ed}{\end{document}}

\newcommand{\np}{\newpage}

\newcommand{\bi}{\begin{itemize}}
\newcommand{\ei}{\end{itemize}}

\newcommand{\bce}{\begin{center}}
\newcommand{\ece}{\end{center}}

\newcommand{\RE}{\,{\rm Re}}
\newcommand{\IM}{\,{\rm Im}}

\newcommand{\sF}{\mathscr{F}}

\begin{document}

\title{Spectral Singularities and a New Method of Generating~Tunable Lasers}

\author{Ali~Mostafazadeh\thanks{E-mail address:
amostafazadeh@ku.edu.tr, Phone: +90 212 338 1462, Fax: +90 212 338
1559}
\\
Department of Mathematics, Ko\c{c} University, \\ 34450 Sar{\i}yer,
Istanbul, Turkey}

\date{ }
\maketitle

\begin{abstract}

We use a simple setup based on an infinite planar slab gain medium
with no mirrors to explore the possibility of realizing a recently
discovered resonance effect related to the mathematical concept of
spectral singularity. In particular we determine the range of the
gain coefficient $g$ and the width $L$ of the gain region required
to achieve this resonance effect. We outline a method that allows
for amplifying waves of desired wavelength by adjusting the gain
coefficient (pumping intensity). We expect this method to have
important practical applications in building tunable lasers.

\hspace{6cm}{Pacs numbers: 03.65.Nk,  42.25.Bs, 42.60.Da, 24.30.Gd}



\end{abstract}

\maketitle

\section{Introduction}

\begin{figure}
    \begin{center}
    \includegraphics[scale=.75,clip]{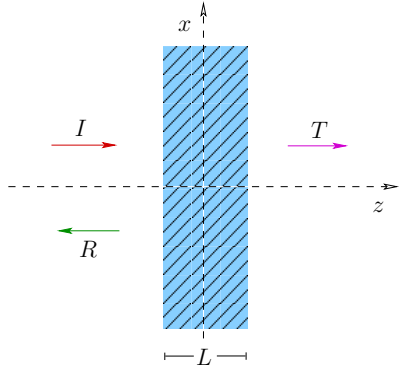}
    \caption{(Color online) Cross section of a planar gain medium (dashed
    region) in the $x$-$z$ plane. Arrows labeled by $I$, $R$, and $T$
    represent the incident, reflected, and transmitted waves.\label{fig1}}
    \end{center}
    \end{figure}

Consider an infinite planar slab gain medium that is aligned along
the $x$-axis, as shown in Figure~\ref{fig1}, and a linearly
polarized monochromatic electromagnetic (EM) wave traveling along
the $z$-axis: $\vec E(z,t)=E\:e^{i(kz-\omega t)}\hat e_x$, where $E$
is a constant and $\hat e_x$ stands for the unit vector pointing
along the positive $x$-axis.
It is easy to show that while traveling through the gain medium the
wave is amplified by a factor of $e^{gL}$, where $g$ is the gain
coefficient and $L$ is the width of the gain medium. For a
homogeneous and isotropic gain medium that is characterized by a
complex refractive index $\mathfrak{n}$, the gain coefficient is
related to the wavelength $\lambda:=2\pi c/\omega$ of the incident
wave and the imaginary part $\kappa$ of $\mathfrak{n}$ (also known
as the extinction factor) according to \cite{silfvast}
    \be
    g=-\frac{4\pi\kappa}{\lambda}.
    \label{gain-coeff}
    \ee
In practice, one usually places the gain medium between two mirrors
(also aligned along the $x$-axis) to produce a (Fabry-Perot)
resonator. In this way one effectively extends the length of the
path of the wave through the gain medium and achieves a much larger
amplification of the wave for the resonance frequencies. In this
article, we will show that it is possible to achieve a similar
amplification effect without using any mirrors provided that we
choose the width of the gain medium and the value of the gain
coefficient in such a way that the system supports a spectral
singularity. This is in effect a method of constructing lasers that
as, we will show, allows for adjusting the frequency of the emitted
radiation by changing the gain coefficient (pumping intensity.)

Suppose that the gain medium is obtained by doping a host medium of
refraction index $n_0$ and that it is modeled by a two-level atomic
system with lower and upper level population densities $N_l$ and
$N_u$, resonance frequency $\omega_0$, and damping coefficient
$\gamma$. Then the complex permittivity $\varepsilon=\fn^2$ of the
system has the form $\varepsilon(z)=\varepsilon_0[1-\hat v(z)]$,
where $\varepsilon_0$ is the permittivity of the vacuum,
    \be
    \hat v(z):=\left\{\begin{array}{ccc}
    \hat\fz &{\rm for}& |z|<\alpha,\\
    0&{\rm for}& |z|\geq\alpha,\end{array}\right.~~~~~
    \hat\fz:=1-n_0^2+
    \frac{\omega_p^2}{\omega^2-\omega_0^2+i\gamma\,\omega} ,
    \nn
    \ee
$\alpha:=L/2$, $\omega_p^2:=(N_l-N_u)e^2/(m\varepsilon_0)$, and $e$
and $m$ are electron's charge and mass, respectively
\cite{silfvast,yariv-yeh}.

We can easily construct the following solution of Maxwell's
equations for the above system.
    \be
    \vec E(z,t)=E\:e^{-i\omega t}\psi(z)\hat e_x,~~~
    \vec B(z,t)=-i \omega^{-1}E\:e^{-i\omega t}\psi'(z)\hat e_y,
    \nn
    \ee
where $\hat e_y$ is the unit vector along the positive $y$-axis, and
$\psi$ is a continuously differentiable solution of the
time-independent Schr\"odinger equation,
    \be
    -\psi''(z)+v(z)\psi(z)=k^2\psi(z),
    \label{sch-eq}
    \ee
for the complex barrier potential $v(z):=k^2\hat v(z)$.

It turns out that the Schr\"odinger operator for a complex barrier
potential may not have a complete set of eigenfunctions \cite{p90}.
This is related to the presence of what mathematicians call a
spectral singularity \cite{ss-math}. Physically, spectral
singularities correspond to the energies $k^2$ at which both the
left and right reflection and transmission coefficients diverge
\cite{prl-09,longhi}. As a result, they define scattering solutions
of (\ref{sch-eq}) that behave exactly like resonances. These states
differ from ordinary resonances, because they have real and positive
energies. Because for a resonance, the imaginary part of the energy
is interpreted as its width, spectral singularities may be viewed as
defining resonances with a zero width \cite{prl-09,zafar-09}.

For the system considered here, infinite reflection and transmission
coefficients mean infinite amplification of an incident EM wave of
finite intensity. Obviously, the presence of an intense radiation
field alters the properties of the gain medium and makes our simple
model unrealistic. However, as a consequence of this amplification
effect and the presence of the background noise, the system emits EM
waves of finite intensity at the frequency of the spectral
singularity. This is a special lasing effect.

In an optical cavity laser the radiation is confined by a set of
mirrors so that it undergoes multiple internal reflections necessary
for satisfying the laser threshold condition. In contrast, for the
system we consider the amplification effect stems from intrinsic
internal reflections that are only present for particular values of
the parameters of the system ($L$ and $g$) and a single frequency
associated with these values.

In \cite{prl-09,p90} we consider a possible realization of this
phenomenon using two different waveguide systems. Here we explore
its consequences for the simpler system of an infinite planar gain
medium which is more convenient to analyze. Unlike for the systems
considered in \cite{prl-09,p90}, here we parameterize the location
of spectral singularities using the gain coefficient $g$ which is an
easily adjustable quantity in practice. Furthermore, we show that by
changing the value of $g$ we can realize the spectral singularity
related resonance effect at wavelengths that differ from the
resonance wavelength of the gain medium. This may be viewed as a
novel method of generating amplified waves of desired wavelength by
adjusting the pumping intensity.

In the remainder of this section, we derive a formula that expresses
$\omega_p^2$ in terms of $g$.

Denoting the real part of the complex refractive index $\fn$ by
$\eta$, so that $\fn=\eta+i\kappa$, and using
$\fn^2=\varepsilon=\varepsilon_0[1-\hat v(z)]$, we have for
$|z|<\alpha$:
    \be
    \eta^2-\kappa^2=n_0^2-\frac{\omega_p^2(\omega^2-\omega_0^2)}{
    (\omega^2-\omega_0^2)^2+\gamma^2\omega^2},~~~~~~~~~
    2\eta\kappa=\frac{\omega_p^2\gamma\,\omega}{
    (\omega^2-\omega_0^2)^2+\gamma^2\omega^2}.
     \label{e1}
    \ee
Setting $\omega=\omega_0$ in (\ref{e1}) and solving for $\omega_p^2$
yields $\omega_p^2=2\kappa\gamma\omega_0\sqrt{n_0^2+\kappa^2}$.
Combining this equation with (\ref{gain-coeff}) and using
$\lambda=2\pi c/\omega$, we find
    \be
    \omega_p^2=-\frac{c\gamma\,\omega_0
    g}{\omega}\sqrt{n_0^2+\frac{c^2g^2}{4\omega^2}}.
    \label{omega-p=}
    \ee

\section{Spectral Singularities of Complex Barrier Potential}

The solution of the Schr\"odinger equation (\ref{sch-eq}) is
straightforward. The following is a set of eigenfunctions of the
complex barrier potential $v(x)$.
    \be
    \psi_{k,\fa}(x)=\left\{\begin{array}{ccc}
    A_-^\fa e^{i k x}+B_-^\fa e^{-i k x} & {\rm for} & x\leq
    -\alpha\\
    A_0^\fa e^{i w k x}+B_0^\fa e^{-i w k x}& {\rm for} & |x|<
    \alpha\\
    A_+^\fa e^{i k x}+B_+^\fa e^{-i k x} & {\rm for} & x\geq
    \alpha
    \end{array}\right.
    \label{eg-fu}
    \ee
where $k\in\R^+$, $\fa\in\{1,2\}$ is a degeneracy label, $A_-^\fa$
and $B_-^\fa$ are free complex coefficients, $w:=\sqrt{1-\hat\fz}$,
and $A_0^\fa, B_0^\fa, A_+^\fa$ and $B_+^\fa$ are complex
coefficients related to $A_-^\fa$ and $B_-^\fa$. We can express this
relationship most conveniently as $\vec C^\fa_0=\underline{L}\:\vec
C^\fa_-$ and $\vec C^\fa_+=\underline{M}\:\vec C^\fa_-$, where
    \bea
    &&\vec C^\fa_0:=\left(\begin{array}{c}
    A_0^\fa\\B_0^\fa\end{array}\right),~~~~
    \vec C^\fa_\pm:=\left(\begin{array}{c}
    A_\pm^\fa\\B_\pm^\fa\end{array}\right),\nn\\
    &&
    \underline{L}:=\frac{1}{2w}\left(\begin{array}{cc}
    e^{i\alpha k(w-1)}(w+1) & e^{i\alpha k(w+1)}(w-1)\\
    e^{-i\alpha k(w+1)}(w-1) & e^{-i\alpha k(w-1)}(w+1)
    \end{array}\right),
    \nn
    \eea
$\underline{M}$ is the transfer matrix:
    {\small\be
     \underline{M}:=\frac{1}{4w}\left(\begin{array}{cc}
    e^{-2i\alpha k}f(w,-\alpha k) & 2i(w^2-1)\sin(2w\alpha k)\\
    -2i(w^2-1)\sin(2w\alpha k)& e^{2i\alpha k}f(w,\alpha k)
    \end{array}\right),
    \nn
    \ee}%
and $f(z_1,z_2):=e^{-2iz_1z_2}(1+z_1)^2-e^{2iz_1z_2}(1-z_1)^2$ for
all $z_1,z_2\in\C$. The spectral singularities are the energy values
$k^2$ for which $M_{22}=0$, i.e., the real $k$ values for which
    \be
    f(w,\alpha k)=0.
    \label{f=0}
    \ee
In Ref.~\cite{p90}, we have provided a detailed analysis of this
equation. Here we give the final result.

\begin{figure}[t]
\begin{center}
\includegraphics[scale=.09,clip]{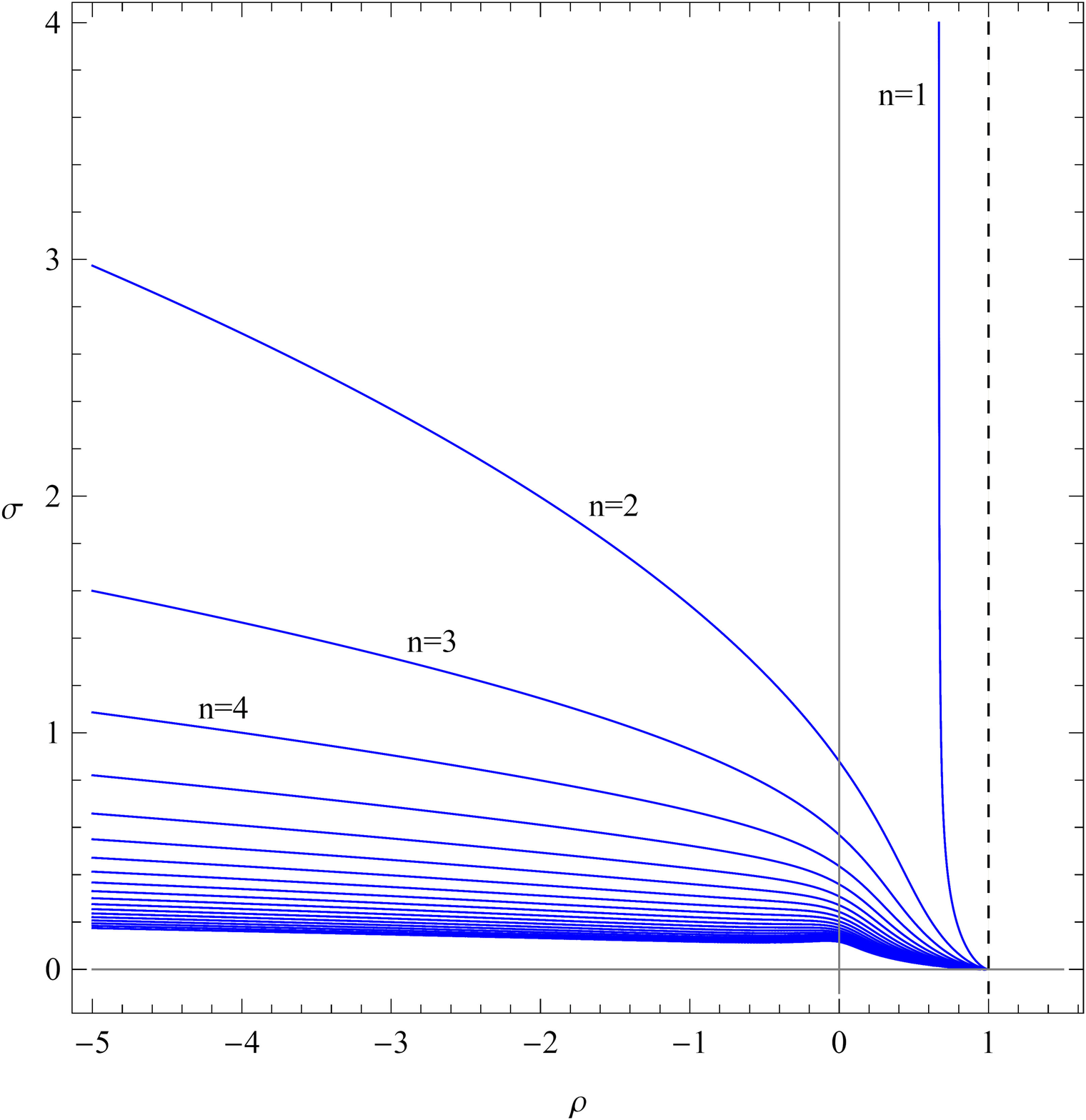}
{\caption{(Color online) Graphs of the curves in the $\rho$-$\sigma$
plane along which there is a spectral singularity. The unmarked
curves correspond to $n=5,6,7,\cdots,20$ from top to bottom,
respectively. As $n\to\infty$ the curve labeled by $n$ tends to the
negative $\rho$-axis. The dashed line corresponds to $\rho=1$. The
solid gray lines are the coordinate axes. \label{fig2}}}
\end{center}
\end{figure}

First, we introduce the parameters: $\rho:=\RE(\hat\fz)=\RE(1-w^2)$,
$\sigma:=\IM(\hat\fz)=\IM(1-w^2)$, and the functions:
    \bea
    g_1(\rho,\sigma)&:=&\cos^{-1}\left[
    \frac{1-\sqrt{(1-\rho)^2+\sigma^2}}{\sqrt{\rho^2+\sigma^2}}\right],
    \label{g1}\\
    g_2(\rho,\sigma)&:=&1-\rho+\sqrt{(1-\rho)^2+\sigma^2},
    \label{g2}
    \eea
where $\cos^{-1}$ stands for the principal value of the inverse of
$\cos$; in particular, $\cos^{-1}(x)\in[0,\pi]$ for all
$x\in[-1,1]$. Then (\ref{f=0}) has a solution for a real $k$ if and
only if $\sigma>0$, $\rho<1$, and for some positive integer $n$,
    \be
    \sinh^2\left\{\big[\pi n-g_1(\rho,\sigma)\big]
    \sqrt{1-\frac{2(1-\rho)}{g_2(\rho,\sigma)}}\right\}=
    \frac{2g_2(\rho,\sigma)}{{\rho^2+\sigma^2}}.
    \label{F=0}
    \ee
This equation determines an infinite family of curves (labeled by
$n$) in the $\rho$-$\sigma$ plane along. See Figure~\ref{fig2}.
For the $\rho$ and $\sigma$ values corresponding to a point on one
of these curves the spectrum of the complex barrier potential $v(x)$
includes a spectral singularity. Furthermore, the wave number
corresponding to this spectral singularity is given by \cite{p90}:
    \be
    k=\frac{\pi n-g_1(\rho,\sigma)}{\alpha\sqrt{2
    g_2(\rho,\sigma)}}.
    \label{k=}
    \ee

\section{Realizing a Spectral Singularity Using a Solid State Gain
Medium}

\begin{figure}[t]
\begin{center}
\includegraphics[scale=.9,clip]{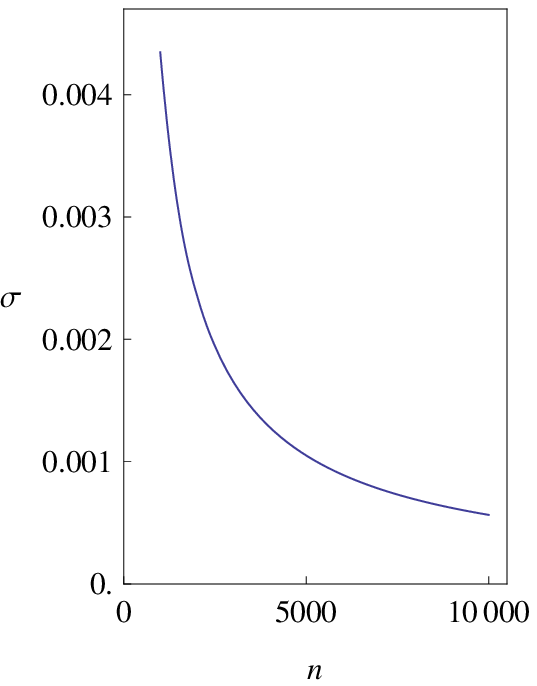}~~~~~~
\includegraphics[scale=.96,clip]{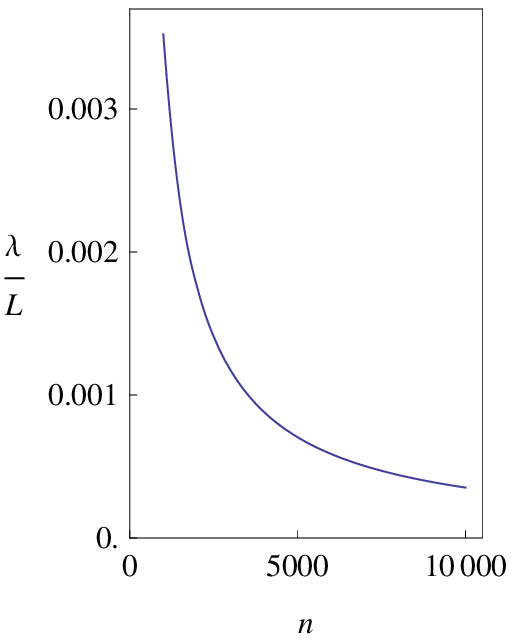}
{\caption{(Color online) Graphs of $\sigma$ (on the left) and
$\lambda/L$ (on the right) as functions of $n$ for
$\lambda=\lambda_0$ and $n_0=1.76$.\label{fig3}}}
\end{center}
\end{figure}

For the particular problem we are considering,
    \be
    \rho=1-n_0^2+
    \frac{\omega_p^2(\omega^2-\omega_0^2)}{(\omega^2-\omega_0^2)^2+
    \gamma^2\omega^2},~~~~
    \sigma=\frac{-\omega_p^2\gamma\,\omega}{(\omega^2-\omega_0^2)^2+
    \gamma^2\omega^2}.
    \label{rh-si=}
    \ee
In order to gain an insight in the condition of the occurrence of
spectral singularities, we first consider the possibility of
creating a spectral singularity at the resonance frequency:
$\omega=\omega_0$. Then Eqs.~(\ref{rh-si=}) simplify considerably,
and in view of (\ref{omega-p=}) we find
    \be
    \rho=1-n_0^2,~~~~~\sigma=\frac{\lambda_0g}{2\pi}
    \sqrt{n_0^2+\left(\frac{\lambda_og}{4\pi}\right)^2},
    \label{reso}
    \ee
where $\lambda_0:=2\pi c/\omega_0$. For a very wide range of
practical situations $\lambda_0g\ll 2\pi$. This together with
(\ref{reso}) imply
    \be
    \sigma\approx \frac{\lambda_ogn_0}{2\pi}\ll 1.
    \label{si-0}
    \ee
As seen from Figure~\ref{fig2}, this relation suggests that we can
tune the parameters of the system to produce a spectral
singularities for large values of $n$. Figure~\ref{fig3} shows the
graphs of $\sigma$ and $\lambda/L$ as a function of $n$ for the case
that $n_0=1.76$, $\lambda=\lambda_0$, and  $1000\leq n\leq 10000$.
For this range of values of $n$, both of these functions turn out to
have the same general (monotonically decreasing) behavior. This is
also true for other typical values of $n_0$.
Using the values $n=1000,1500,2000,2500,\cdots, 10000$ we could
establish that for this range of values of $n$, both $\sigma$ and
$\lambda/L$ are approximately inversely proportional to $n$;
$\sigma\approx (4.71\pm0.60)/n$ and $\lambda/L\approx (3.19\pm
0.33)/n$.

\begin{figure}[t]
    \begin{center}
    \includegraphics[scale=.9,clip]{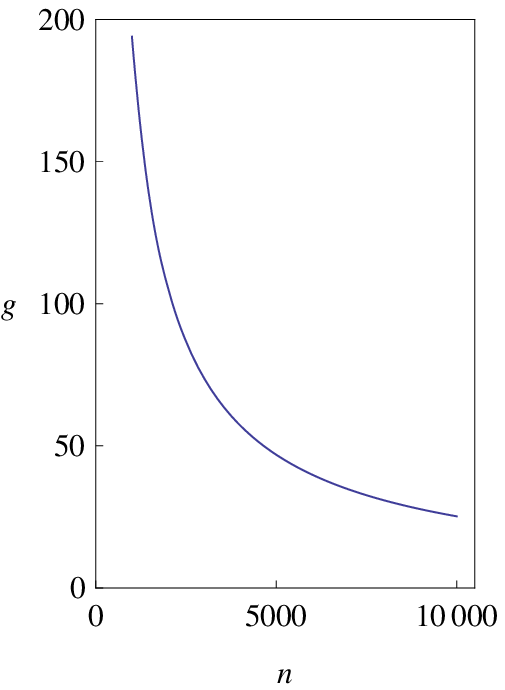}~~~~~~
    \includegraphics[scale=.91,clip]{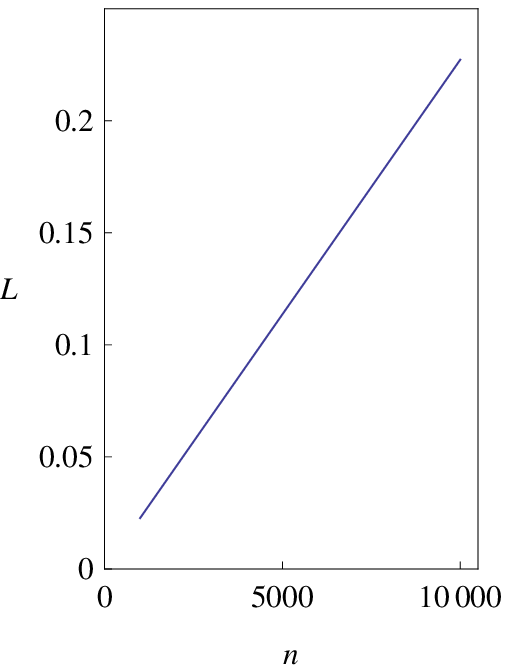}
{\caption{(Color online) Graphs of the gain coefficient $g$ in
cm$^{-1}$ (on the left) and the length of the gain region
$L=2\alpha$ in cm (on the right) as functions of $n$ for
$\lambda_0=800$~nm and $n_0=1.76$.\label{fig4}}}
\end{center}
\end{figure}

Let us now take $\lambda_0=800$~nm for the resonance wavelength and
again set $n_0=1.76$. Then as we increase the value of $n$ from
$1000$ to $10000$, the gain coefficient $g$ (respectively the width
of the gain region $L=2\alpha$) that would correspond to a spectral
singularity decreases from 193.98/cm to 25.19/cm (increases from
0.22705 mm to 2.2725 mm). Figure~\ref{fig4} shows the graphs of $g$
and $L$ as a function of $n$.
For $1000\leq n\leq 10000$, $g$ is a monotonically decreasing
function of $n$ while $L$ is monotonically increasing function of
$n$. Using the values of $g$ and $L$ for
$n=1000,1500,2000,2500,\cdots, 10000$, we find that
    \be
     g\approx (2.10\pm 0.27)\times 10^5/n~~{\rm cm}^{-1},~~~~
    L\approx (2.06\pm 0.21)\times 10^{-5}
    \,n~~{\rm cm},~~~~g\,L\approx 4.78\pm 0.61.
    \label{approx}
    \ee

The range of gain coefficients that we obtain for $1000\leq n\leq
10000$ is actually quite high for a typical solid state gain media
such as a Titanium Sapphire crystal with $n_0=1.76$ and
$\lambda_0=800~{\rm nm}$, \cite{silfvast}. To obtain a spectral
singularity for a typical gain coefficient for this crystal, we need
to take a much larger $n$. This in turn requires a much larger gain
region. For example, for $g=0.2~{\rm cm}^{-1}$, (\ref{approx})
suggests that $n\gtrsim 10^6$ and $L\gtrsim 20~{\rm cm}$. A direct
calculation shows that setting $n=1.945\times 10^6$ we find a
spectral singularity for $g=0.200~{\rm cm}^{-1}$ and $L=44.2~{\rm
cm}$. This is an unrealistically large number for the size of a
solid state gain medium with a uniform gain coefficient. Therefore,
an experimental study of spectral-singularity-related resonance
effect that uses the simple setup we outlined above requires a gain
medium with relatively high gain coefficient. Possible alternatives
are the gain media used in dye lasers or the semiconductor diode
lasers.

\section{Realizing a Spectral Singularity Using a Semiconductor Gain
Medium}

In this section we consider the theoretical plausibility of
achieving a spectral singularity related resonance effect using a
semiconductor gain medium. Here we disregard the experimental
difficulties of realizing our simple setup.

Consider a gain medium used in building a typical diode laser
\cite{silfvast} with $n_0=3.4$ and
    \be
    \lambda_0=0.35-24~\mu{\rm m},~~~
    \gamma=1.57-6.28\times 10^{13}~{\rm
    Hz},~~~
    g=(1-10)\times 10^4~{\rm m}^{-1},~~~ L=200-500~\mu{\rm
    m}.
    \label{range-diode}
    \ee
To be specific we first examine the case that
    \be
    \lambda_0=1.5~\mu{\rm m},~~~\gamma=3\times 10^{13}~{\rm
    Hz},~~~
    g=5\times 10^4~{\rm m}^{-1},~~~L=350~\mu{\rm m},
    \label{diode}
    \ee
$gL=17.5$, and $\lambda_0g=7.5\times 10^{-2}\ll 1$. We note that the
above value of $g$ is the maximum value at which the diode laser
functions. In other words, we are allowed to adjust gain coefficient
to any value between 0 and $5\times 10^4~{\rm m}^{-1}$.

In view of (\ref{reso}) and $\lambda_0g\ll 1$, we can use
(\ref{si-0}) to estimate the value of $\sigma$. This gives
$\sigma\approx 0.04$. We find by inspection that to maintain this
value we should take $n\approx 110$. It turns out that for $n=111$,
the necessary gain coefficient is given by $g=4.9959\times 10^4~{\rm
m}^{-1}$. This in turn corresponds to $L= 24.26~\mu{\rm m}$ which
lies outside the allowed range (\ref{range-diode}).

Examining the value of the product $gL$ for $n$ between 110 and 114
we see that its average value and standard deviation are
respectively $1.21$ and $1.8\times 10^{-7}$. Hence at least for this
range of values of $n$, $gL$ is essentially independent of $n$.
Furthermore, $g$ (respectively $L$) is a decreasing (increasing)
function of $n$. This shows that to obtain realistic values of $g$
and $L$ as given in (\ref{range-diode}) we should consider larger
values for $n$.

Suppose that we wish to use a sample with $L\approx 350 \mu{\rm m}$.
Then given $gL\approx 1.21$, we estimate that $g\approx 3500 {\rm
m}^{-1}$, and in view of (\ref{si-0}), $\sigma\approx 0.0028$. This
turns out to correspond to $n\approx 1600$. Direct calculation
confirms the existence of a spectral singularity with $n=1588$,
$g=3462.91~{\rm m}^{-1}$, and $L=350.074~\mu{\rm m}$. We have also
checked that for $n=1585-1589$, the average and standard deviation
of $gL$ are respectively given by $1.21$ and $2.88\times 10^{-9}$,
i.e., at least for large values of $n$, $gL$ is essentially
independent of $n$. In particular, $g\approx 1.21/L$.

\section{Adjusting the Wavelength of the Emitted Wave by Changing
the Pumping Intensity}

Consider the following practically important scenario. Suppose that
we fix the width of the gain region $L$ so that there is a spectral
singularity at the resonance wavelength $\lambda_0$ for a particular
value of the gain coefficient $g_\star$ and the label $n$ that we
denote by $n_\star$. If we vary the value of the gain coefficient
$g$ in the vicinity of $g_\star$, we can adjust it to create a new
spectral singularity for some $\lambda$ and $n$ in the vicinity of
$\lambda_0$ and $n_\star$. This is an important observation, for we
can use it as a method of generating amplified waves of wavelength
close to $\lambda_0$ by adjusting the value of the gain coefficient,
i.e., carefully tuning the pumping intensity. As far as we know,
this is the first instance of a method that uses the change in
pumping intensity to fine tune the wavelength of the emitted wave.
This possibility marks an important distinction between the spectral
singularity related resonances and the usual resonances that one
encounters in the usual optical resonators.

The mathematical implementation of this idea involves the following
steps. First, we express (\ref{F=0}) and (\ref{k=}) as equations in
the three variables $n$, $g$, and $\lambda$. Next, we use (\ref{k=})
to express $n$ in terms of $g$ and $\lambda$. We can summarize the
result as $n=\sF_n(g,\lambda)$. Finally, we substitute this
expression in (\ref{F=0}) to obtain an equation of the form
$\sF(g,\lambda)=0$. The spectral singularities are located on the
curve $C$ defined by this equation in the $g$-$\lambda$ plane, but
not all the points on this curve correspond to a spectral
singularity. This is because by eliminating the variable $n$ in our
calculations we have ignored the fact that $n$ takes integer values
only. We can graphically impose this condition by plotting the
curves $C_n$ defined by $\sF_n(g,\lambda)=n$ for each $n$ in the
vicinity of $n_\star$. The intersection of $C$ with $C_n$ gives the
points in the $g$-$\lambda$ plane that correspond to spectral
singularities.

\begin{figure}[t]
    \begin{center}
    \includegraphics[scale=.9,clip]{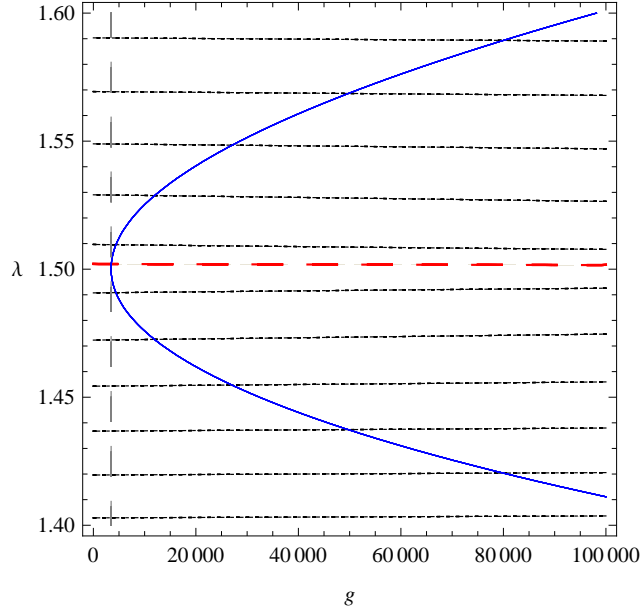}~~~~~~
{\caption{(Color online) Graphs of the curves $C$ (solid, blue
curve) and $C_n$ (the dotted black curves) for
$n=1500,1520,1540,\cdots, 1700$ (from top to bottom) in the
$g$-$\lambda$ plain. $g$ and $\lambda$ values are given in units of
$\mu{\rm m}$ and ${\rm m}^{-1}$, respectively. The vertical dashed
(gray) line is the plot of $g=g_\star=3463~{\rm m}^{-1}$. The thick
dashed red curve is the graph of $C_{1588}$. The spectral
singularities appear in the intersection points of $C$ and
$C_n$.\label{fig5}}}
\end{center}
\end{figure}

As a concrete example consider a diode laser gain medium with
    \be
    n_0=3.4,~~~\lambda_0=1.5~\mu{\rm m},~~~\gamma=3\times 10^{13}~{\rm
    Hz},~~~
    g_\star=3463~{\rm m}^{-1},~~~~~L=350.1~\mu{\rm m},
    \label{diode2}
    \ee
that corresponds to setting $n_\star=1588$. Figure~\ref{fig5} shows
the graphs of the curves $C$ and $C_n$ with
$n=1500,1520,1540,\cdots, 1700$ and $n=1588$. For this range of
values of $n$ the gain coefficient is within the range given in
(\ref{range-diode}).
As seen from this figure to amplify a wave of wavelength
$\lambda\neq\lambda_\star=1.5~\mu{\rm m}$, we need to increase the
gain coefficient. For a particular discrete set of values of
$g>g_\star$ we obtain spectral singularities at certain wavelengths
grouped in pairs with very small difference and ranging between
$1.4$ and $1.6~\mu{\rm m}$. Table~\ref{table1} gives the wavelength
of the spectral singularities obtained by changing $g$ in the range
$60000$-$70000~{\rm m}^{-1}$. According to this table, this allows
for amplifying waves with wavelength in the ranges
1.4263-1.43051~$\mu{\rm m}$ and 1.5768-1.5820 $\mu{\rm m}$.
    \begin{table}
    \begin{center}\vspace{.3cm}
    {\begin{tabular}{|c|c|c||c|c|c|}
    \hline
    $g~({\rm m}^{-1})$ & $\lambda~(\mu{\rm m})$ & $n$ &
    $g~({\rm m}^{-1})$ & $\lambda~(\mu{\rm m})$ & $n$\\
    \hline
    \hline
    60929 & 1.4305 & 1668 & 65440 & 1.4280 & 1671\\
    \hline
    60966 & 1.5768 & 1512 & 65475 & 1.5710 & 1509 \\
    \hline \hline

    62414 & 1.4297 & 1669 &  66982 & 1.4271 & 1672\\
    \hline
    62450 & 1.5779 & 1511 & 67015 & 1.5810 & 1508\\
    \hline \hline

    63918 & 1.4288 & 1670 & 68542 & 1.4263 & 1673\\
    \hline
    63953 & 1.5789 & 1510 &  68575 & 1.5820 & 1507\\
    \hline
    \end{tabular}}
    \caption{The spectral singularities for a semiconductor gain
    medium with $n_0=3.4$, $\lambda_0=1.5~\mu{\rm m}$,
    $\gamma=3\times 10^{13}~{\rm  Hz}$, $L=350.1~\mu{\rm m}$,
    obtained at various wavelengths as one changes the gain
    coefficient $g$ in the range $60000$-$70000~{\rm m}^{-1}$.\label{table1}}
    \end{center}
    \end{table}

\section{Concluding Remarks}

In this article, we examined the possibility of realizing the
spectral-singularity-related resonance (lasing) effect by using the
gain coefficient to parameterize the problem. We showed that an
experimental verification of this effect that involves a typical
solid state gain medium is unrealistic and that we needed a gain
medium of relatively high gain coefficient. We therefore explored
the possibility of using a semiconductor gain medium.

More importantly, we proposed a method that allows for amplifying
waves of wavelength different from the resonance wavelength of the
gain medium by simply adjusting the gain coefficient. This is an
interesting observation as one can change the gain coefficient by
changing the pumping intensity. The toy model we used to examine the
above issues is probably too simple for  modeling a real experiment.
Here we considered this system because it revealed some of the
remarkable consequences of the spectral singularity-related
resonance effect.\vspace{.3cm}

\noindent {\em Acknowledgments:} I wish to thank Aref Mostafazadeh
for illuminating discussions. This work has been supported by the
Scientific and Technological Research Council of Turkey
(T\"UB\.{I}TAK) in the framework of the project no: 108T009, and by
the Turkish Academy of Sciences (T\"UBA).

\np

\ed